\documentclass[american,aps, preprint, showpacs, superscriptaddress]{revtex4-1}
\usepackage[T1]{fontenc}
\usepackage[utf8]{inputenc}
\usepackage{color}
\usepackage{babel}
\usepackage{array}
\usepackage{booktabs}
\usepackage{textcomp} 
\usepackage{amsmath, physics}
\usepackage{amssymb}
\usepackage{graphicx}
\usepackage{lineno}
\usepackage[unicode=true,
 bookmarks=true,bookmarksnumbered=false,bookmarksopen=false,
 breaklinks=false,pdfborder={0 0 0},pdfborderstyle={},backref=false,colorlinks=true]
 {hyperref}
\hypersetup{colorlinks=true,linkcolor=blue,citecolor=blue}

\makeatletter

 \@ifundefined{textcolor}{}
 {%
   \definecolor{BLACK}{gray}{0}
   \definecolor{WHITE}{gray}{1}
   \definecolor{RED}{rgb}{1,0,0}
   \definecolor{GREEN}{rgb}{0,1,0}
   \definecolor{BLUE}{rgb}{0,0,1}
   \definecolor{CYAN}{cmyk}{1,0,0,0}
   \definecolor{MAGENTA}{cmyk}{0,1,0,0}
   \definecolor{YELLOW}{cmyk}{0,0,1,0}
 }

\usepackage{comment}
\usepackage{lmodern}
\graphicspath{{images/}}

\makeatother

\begin{document}

\title{Rabi oscillations of dissipative structures effected by out-of-phase parametric drives}

\author{Fernando R. Humire}
\email{fr.humire@gmail.com}
\affiliation{Departamento de F\'isica, Facultad de Ciencias, Universidad de Tarapac\'a, Casilla 7-D Arica, Chile}

\author{Yogesh N. Joglekar}
\affiliation{Department of physics, Indiana University-Purdue University Indianapolis, Indianapolis, Indiana 46202, United States }

\author{M\'onica A. Garc\'ia-~Nustes }
\affiliation{Instituto de F\'{i}sica, Pontificia Universidad Cat\'olica de Valpara\'{i}so,
Casilla 4059, Chile}

\begin{abstract}
Dissipative structures are localized, stable patterns that arise due to the intricate balance among dissipation, dispersion, interaction, and external drive. Their creation and manipulation are of great interest in fields as diverse as optics, magnetism, fluids, and the quantum theory. Here, we report on the emergence of Rabi oscillations of these patterns in a dissipative, nonlinear system subject to two localized, out-of-phase, parametric drives. Their period and amplitude are controlled by the drive size and separation, and can be varied over a wide range.  We show that this system undergoes a transition similar to the prototypical parity-time ($\mathcal{PT}$) symmetry breaking transition, but in contrast to the usual $\mathcal{PT}$-symmetric models, these oscillations are robust against the drive mismatch. By using a lossy, nonlinear dimer model with out-of-phase drives, we analytically explain some of these findings. Our results demonstrate a new way to robustly balance gain and loss in a globally dissipative medium, and manipulate the dissipative structures. 
\end{abstract}
\maketitle

Dissipative structures arise in out-of-equilibrium systems where a continuous energy flow sets in between the system and its surroundings. This paradigm was introduced in 1967 by I. Prigogine to describe a stable response in an out-of-equilibrium  nonlinear system~\cite{Turing1952,Nicolis1977,Castets1990}. The energy is continuously redistributed  between the different parts of the structure, constituting a long-lived form, i.e. self-organization~\cite{AchmedievAnkiewicz2005,Erwin2016}. In other words, dissipative structures maintain an ``inner balance'' between frictional loss and gain from external driving that is robust against large variations in both parameters. Patterned structures~\cite{Descalzi2011} such as Faraday waves and dissipative solitons are two representative examples resulting from such inner balance~\cite{Cross1993,Cross2009}.  We note that interactions and dispersion are instrumental to the formation and stability of such patterns arising from a quiescent equilibrium state. Their robustness, on the other hand, makes the manipulation and steering of these patterns challenging~\cite{Gordillo2011}.

Over the past decade, a new paradigm for balancing gain and loss, called parity-time ($\mathcal{PT}$) symmetric systems, has emerged~\cite{Joglekar2013,El-Ganainy2018}. It is described by non-relativistic Schr\"{o}dingner equation with a complex potential $V(x)\neq V^*(x)$ that is invariant under combined operations of parity, given by $\mathcal{P}:x\rightarrow -x$, and time reversal, given by complex conjugation *. Such potentials represent balanced gain ($\Im V>0$) and loss ($\Im V<0$), and have been engineered in diverse platforms comprising optical waveguides~\cite{Ruter2010}, resonators~\cite{Peng2014,Peng2014a}, acoustics \cite{Zhu2014}, electrical circuits~\cite{Schindler2011,Leon-Montiel2018}, and mechanical oscillators~\cite{Bender2013b}. Non-linear Schr\"{o}dinger equation with $\mathcal{PT}$-symmetric potentials, too, has been extensively investigated \cite{Konotop2016}. In discrete lattices~\cite{Sukhorukov2010,Barashenkov2014,Barashenkov2015,Feijoo2015}, these studies have led to nonlinear stationary modes such as solitons~\cite{Suchkov2011,Alexeeva2012,He2013}, compactons~\cite{Bender2009}, breathers~\cite{Barashenkov2012}, and and non-trivial dynamics~\cite{KevrekidisP.2013}. In the continuum cases, they have led to bright and dark solitons~\cite{Musslimani2008}, breathers~\cite{Barashenkov2012b}, and localized modes. Some of these ideas have been generalized to $\mathcal{PT}$-symmetric deformations of physically relevant, nonlinear equations such as Korteweg-de Vries (KdV)~\cite{Fring2007,Bender2007kdv} or Burgers equation~\cite{Yan2013}. In all of these cases, localized gain and loss are introduced through $\mathcal{PT}$-symmetric potentials, and are always balanced. 

In this paper, we connect the two paradigms, of dissipative structures and $\mathcal{PT}$-symmetric systems, in a nonlinear, out-of-equilibrium system with global dissipation and two localized, out-of-phase parametric drives. When the drives are well-separated, the system is described by two parametrically driven and damped nonlinear Schr\"{o}dinger equations, each supporting a steady-state pattern localized around the drive-site. When the drives are close-by, we show that the localized pattern shows Rabi oscillations between the two drive sites. The amplitude and period of these oscillations are reminiscent a prototypical $\mathcal{PT}$ transition in a dimer model, but, in a stark contrast with the linear-dimer model, we show that the nonlinearity makes this transition robust over a wide range of parameters including drive mismatch. Thus, our results evince a new way to balance of gain and loss in an out-of-equilibrium system and to manipulate resultant dissipative structures. 

\begin{figure}[t]
	\centering
	\includegraphics[width=0.49\columnwidth]{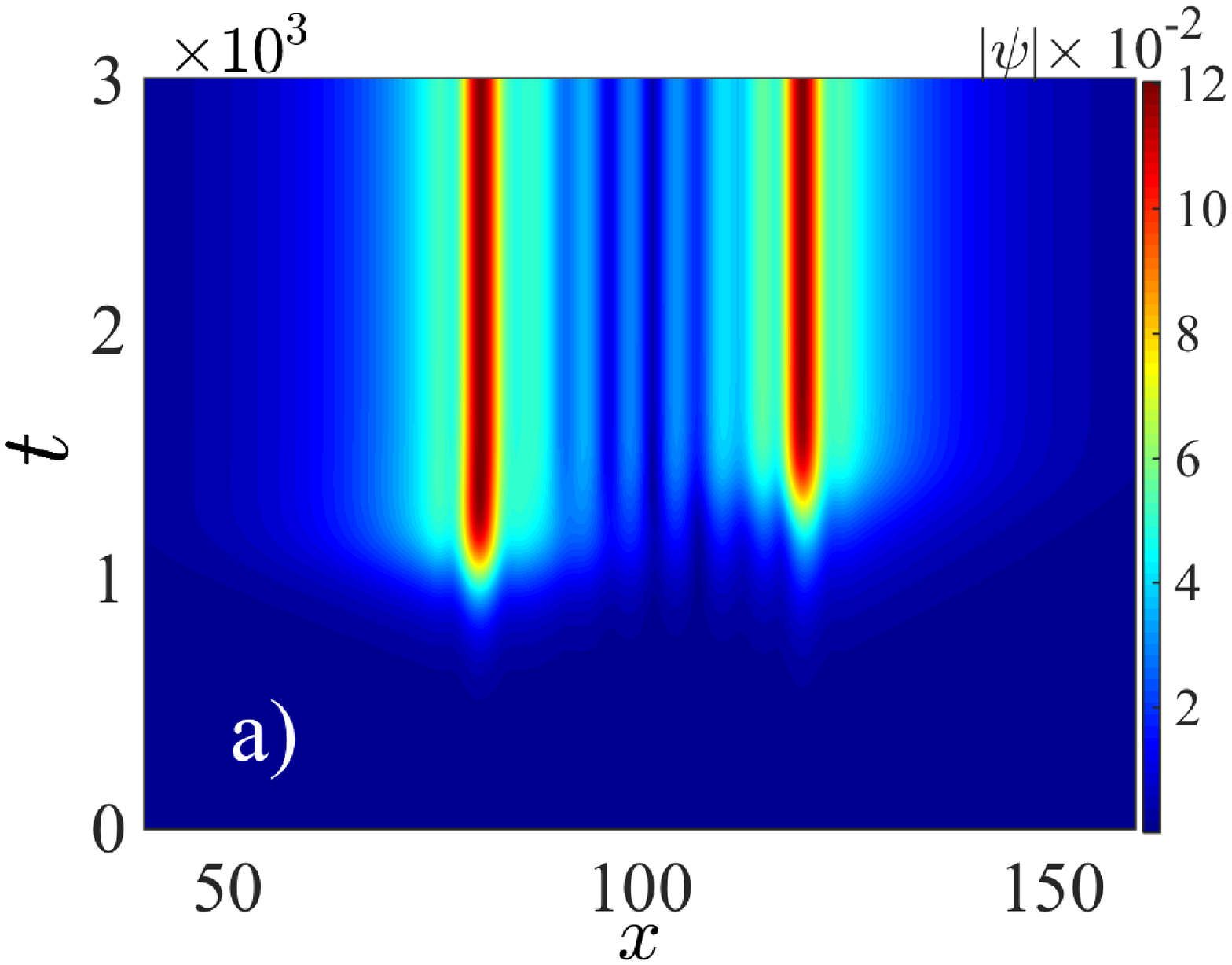}
	\includegraphics[width=0.49\columnwidth]{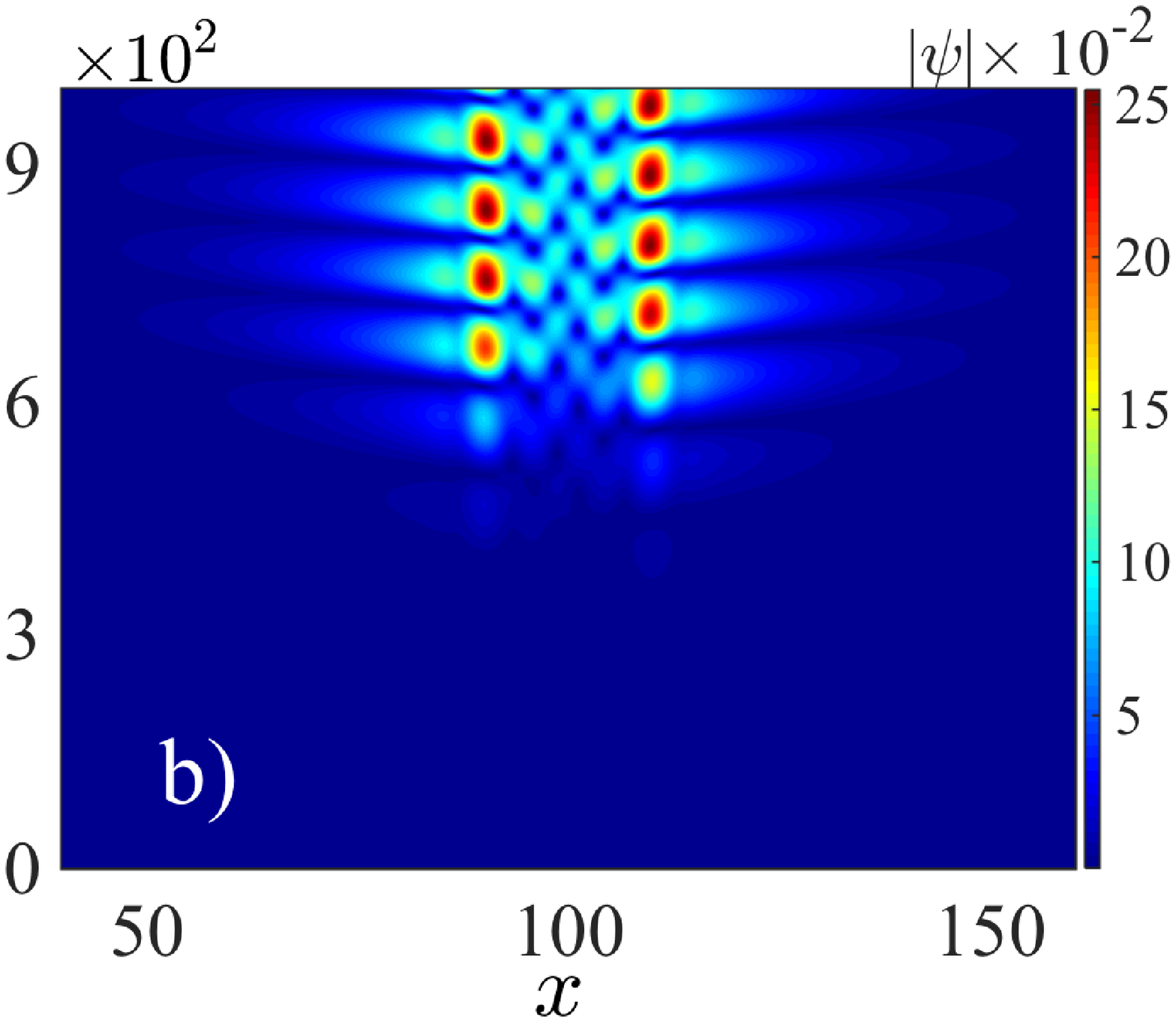}
	\caption{Evolution of the pattern, $|\psi(x,t)|$, with localized, out-of-phase drive separations, and common set of parameters $\mu=0.1,\nu=0.32, \gamma_{\pm}=\gamma_{0}=0.28$, and $\sigma_\pm=\sigma=3$. a) at large separations, $a=39$, two steady-state patterns are established at the locations of the two drives, with interference fringes between the two. b) at smaller separations, $a=20$, a single pattern undergoing Rabi oscillations between the two drive locations emerges due to the cross-talk between the out-of-phase drives. Both solutions emerge from random initial conditions and the system is far removed from the equilibrium quiescent state.}
	\label{Fig:SpatioTemporal}
\end{figure}

\noindent {\it The model.} The parametrically driven and damped NonLinear Schr\"{o}dinger  equation (pdNLSe) describes the amplitude envelope of the oscillations for parametric, extended systems. As its experimental realization, we keep a vertically vibrated shallow tank of water in mind~\cite{Miles1990,Urra2017}. In the dimensionless form, the pdNLSe is given by 
\begin{equation}
\partial_{t}\psi=-i\nu \psi-i|\psi |^{2}\psi-i\partial_{xx}\psi-\mu \psi+\gamma\psi^*
\label{Eq:PPDNLS-1}
\end{equation}
where $\psi(x,t)$ stands for the complex envelope, and $(x,t)$ denote dimensionless space and time respectively. The dimensionless detuning $\nu$ denotes the offset from the parametric resonance frequency, $\mu$ accounts for the global dissipation (energy losses) in the system, i.e. friction, and  $\gamma$ is the amplitude of the parametric driving. For a negative detuning $\nu<0$, the system supports dissipative soliton solutions. When $\nu>0$, a uniform parametric drive with $\gamma>\mu$ and $\gamma^{2}<\nu^{2}+\mu^{2}$, and $\nu\sim\mu\sim\gamma\ll1$, leads to subharmonic patterns with critical wavelength $k_{c}=\sqrt{\nu}$ known as Faraday waves~\cite{Faraday1931,Kumar1996}.  In the following, we only focus on the pattern-forming region, i.e. $\nu>0$. When the parametric drive amplitude $\gamma(x)$ is spatially localized around $x_0$, the formation of patterns localized around $x_0$, instead of Faraday waves, still occurs but the critical threshold $\gamma_m^c$ for each mode $m$ is larger, i.e. $\gamma_m^c>\gamma^c=\mu$.  In other words, to overcome the uniform frictional losses $\mu$, the system requires larger drive amplitude if the injection is localized and the size of the pattern scales as the square-root of the injection-size~\cite{Urra2017}. It follows that when $\gamma(x)$ is a sum of localized, well-separated functions, the result is a superposition of corresponding localized patterns, irrespective of the relative phases of the parametric drives. But what if the localized drives are not well separated? How do the resulting localized patterns interact with each other?  

Motivated by these considerations, we consider a Bi-Gaussian parametric drive with amplitudes $\gamma_\pm$, centers at $x_\pm=L/2\pm a$, and widths $\sigma_\pm$, i.e. 
\begin{eqnarray}
\gamma(x)= \gamma_{+}(x)-\gamma_{-}(x), 
\label{Eq:gaussian_injection}
\end{eqnarray}
where $\gamma_{\pm}(x) = \gamma_{\pm}\exp\left[-(x-x_\pm)^2/2\sigma_\pm^2\right]$ and the system ranges from $0\leq x\leq L$. We note that since each Gaussian drive is not normalized, its net strength, i.e. $\int dx\gamma_\pm (x)$, is proportional to its width $\sigma_\pm$. The negative sign in Eq.(\ref{Eq:gaussian_injection}) denotes the $\pi$-phase delay between the two localized drives. In particular, with equal amplitudes $\gamma_{\pm}=\gamma_0$, and equal widths, $\sigma_{\pm}=\sigma$, Eq.(\ref{Eq:gaussian_injection}) mimics a balanced, spatially separated, energy injection-extraction configuration. We explore the dynamical behavior of resulting localized patterns, through direct numerical simulations of Eq.~\eqref{Eq:PPDNLS-1} with the Bi-Gaussian drive, Eq.(\ref{Eq:gaussian_injection}). With a 400-point wide spatial grid and resolution $\mathrm{d}x=0.5$, we run a simple Runge-Kutta scheme on \eqref{Eq:PPDNLS-1} with time-step $\mathrm{d}t=0.1$ and random initial conditions to obtain the space- and time-dependent $\psi(x,t)$. The Bi-Gaussian function is centered on the grid, and the parameter $2a=|x_{+}-x_{-}|$ controls the separation between the function maxima. 

 \begin{figure*}[t]
	\centering
	\includegraphics[width=0.49\textwidth]{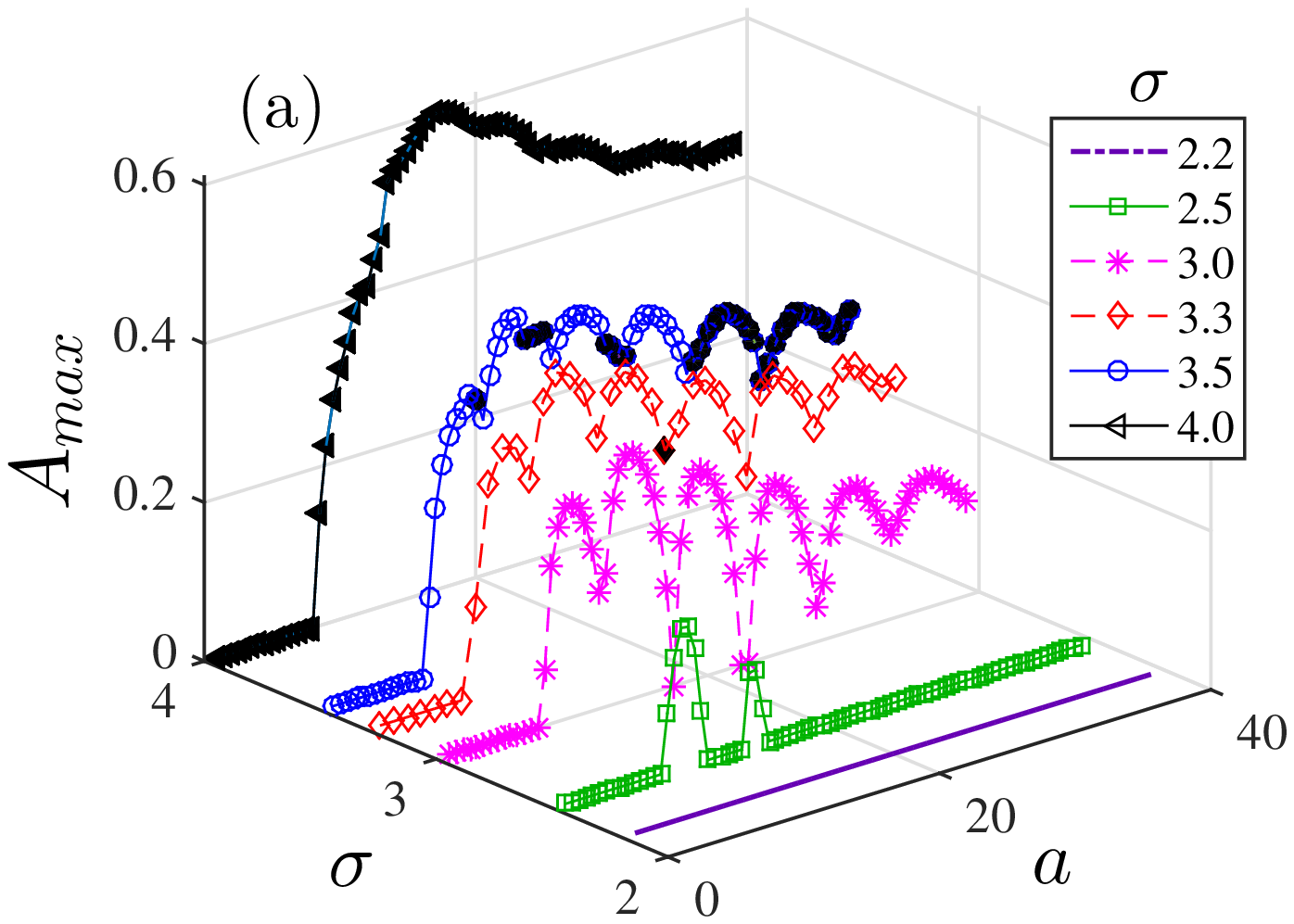}
	\includegraphics[width=0.49\textwidth]{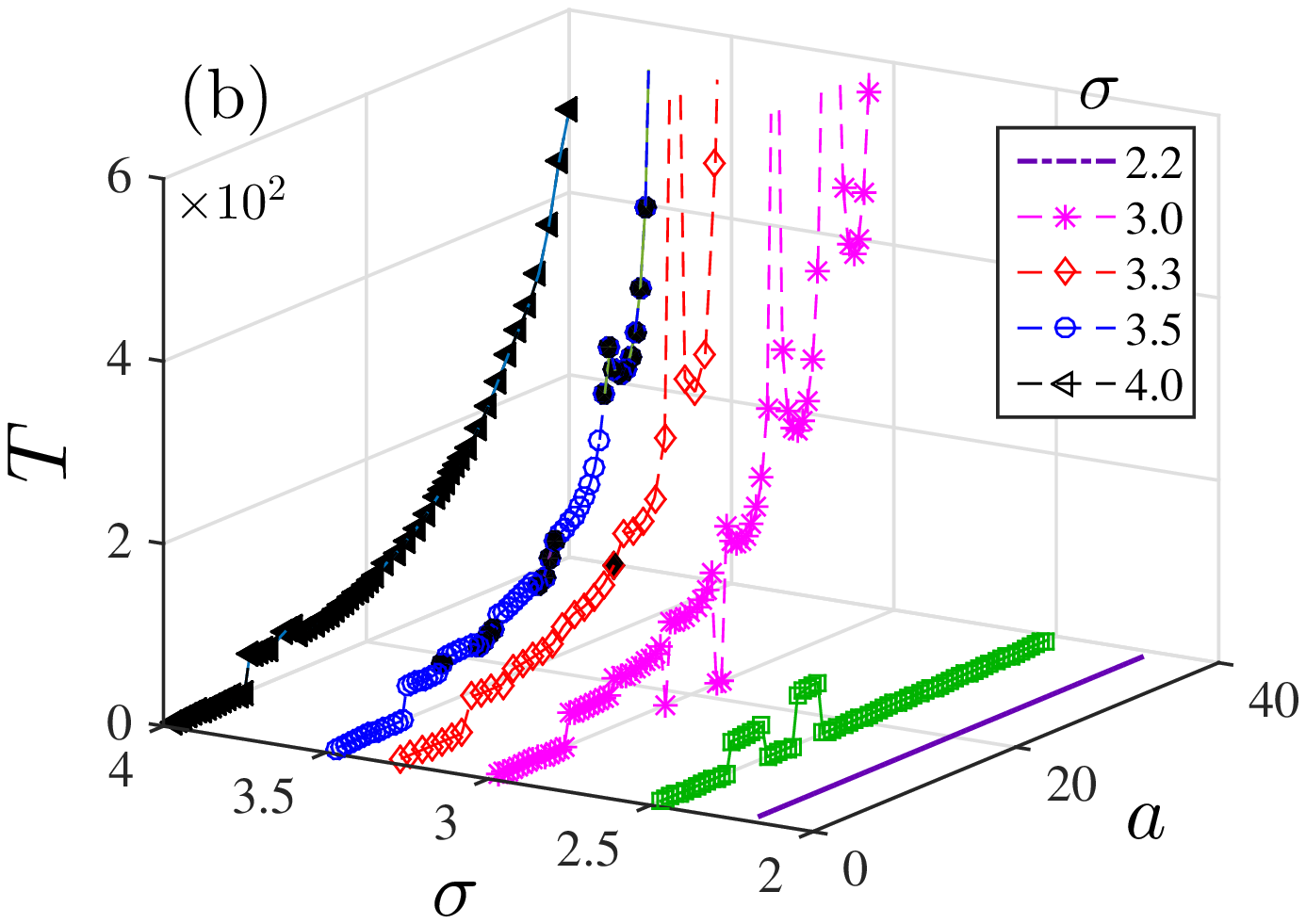} 
	\caption{Characterizing pattern formation and its Rabi oscillations through the pattern amplitude $A_\mathrm{max}(a,\sigma)$ and period $T(a,\sigma)$. The system parameters are $\mu=0.1,\nu=0.32, \gamma_{0}=0.28$. a) for small drive strength or small separation between out-of-phase drives, there is no pattern $A_\mathrm{max}=0$. At large $a$, with sufficiently strong drive, two stable, steady-state patterns emerge, Fig.~\ref{Fig:SpatioTemporal}a, whereas for small $a$, the system has a single pattern undergoing Rabi oscillations between the drive locations. b) The period $T$ of these oscillations diverges at the threshold, and for the most part, decreases monotonically as $a$ is decreased. The steady-state to Rabi-oscillations transition is evocative of the $\mathcal{PT}$-breaking transition in a $\mathcal{PT}$-symmetric dimer with saturating gain.} 
	\label{Fig:Amplitude_Period_sigma_a}
\end{figure*}

Figure~\ref{Fig:SpatioTemporal} shows the resulting $|\psi(x,t)|$ behavior for two different values of $a$ for a given set of $\mu=0.1,\nu=0.32,\gamma_{0}=0.28$, and $\sigma=3$. At a large separation, $a=39$, we see the establishment of two localized, steady-state patterns. It is worth its while to point out that the parametric drive with a positive $\gamma$ leads the formation of the pattern, while the other drive, with negative $\gamma$ has a lag. Since the two drives are not very well separated, the region between them shows oscillatory behavior indicative of the cross-talk between the two patterns. In a sharp contrast, when the separation is small, $a=20$, we see a strikingly different dynamical behavior. Instead of two stationary patterns, the system develops an oscillatory pattern that is reminiscent of Rabi oscillations in a two-level system; i.e. when $|\psi(x,t)|$ is maximum at $x_{+}$, it is vanishingly small at $x_{-}$ and vice versa. We also note that this dynamic pattern is established from random initial conditions faster than its counterpart at large $a$. 

We characterize the oscillating or steady-state patterns by two parameters, the maximum amplitude $A_\mathrm{max}$ and the Rabi oscillation period $T$, and determine their dependence on $(a,\sigma)$, i.e. parameters that characterize the Bi-Gaussian, counter-phase parametric drive. Figure~\ref{Fig:Amplitude_Period_sigma_a} depicts the results of such numerical investigation for the same set of $\mu,\nu$, and $\gamma_0$ values. When the width $\sigma$ is small, the net strength of each drive cannot compensate for the global dissipation $\mu$, and there is no localized pattern. In this regime, where no significant pattern amplitude is detected, i.e. $A_\mathrm{max}=0$, we assign $T=0$ since the question of oscillation period is moot. When the drive strength $\sigma$ is increased, at small $a<a_c\sim 7.9$, there is no steady-state pattern. However, for larger separations $a\geq a_c$ and sufficiently strong drive $\sigma\geq \sigma_c=2.5$, we find a non-monotonic dependence of the pattern amplitude on these two parameters. In particular, the oscillatory dependence of $A_\mathrm{max}$ on $a$ is due to the interference governed by the separation parameter $a$ and the pattern wavelength. Figure~\ref{Fig:Amplitude_Period_sigma_a}b shows the corresponding results for the Rabi oscillation period $T(a,\sigma)$, where $T=0$ is assigned when there is no pattern. For a fixed drive strength $\sigma\geq \sigma_c$, as the drive separation $2a$ is increased, we see that the period increases and diverges at a finite value of $a$. For larger separations, the system has two, steady-state patterns, and the pattern amplitude $A_\mathrm{max}(a,\sigma)$ reaches a constant value that is proportional to $\sqrt{\sigma}$~\cite{Urra2017}. We note that at large $\sigma$, the dynamical behavior of the patterns becomes more exotic, with drifting motion and chaotic behavior; the filled black circles denote their presence. We do not address them further in this work.

Thus, in brief, the pdNLSe with localized, spatially separated, out-of-phase parametric drive has two transitions. The first is the well-known emergence of a localized, steady-state pattern at the drive location. The second is the emergence of Rabi oscillations between these localized patterns, when the parametric drives that generate them are close to each other. Although each pattern - in isolation -- is stabilized by the nonlinearity, global dissipation, and local parametric drive, the motion of these patterns is induced solely by the out-of-phase nature of the two localized drives. The phenomenology seen here, a transition from a steady-state behavior to Rabi oscillations with a period that diverges at the transition, is identical to that of $\mathcal{PT}$-symmetry breaking transition in a $\mathcal{PT}$-symmetric dimer~\cite{Joglekar2013,El-Ganainy2018} with a saturating gain. What is remarkable is that this transition occurs in a globally dissipative, highly nonlinear system that is far removed from its equilibrium quiescent state. 


To further characterize this transition, we consider the behavior of the integrated pattern weight, 
\begin{equation}
\label{eq:weight}
Q(t)=\int_0^L |\psi(x,t)|^2\,dx.
\end{equation}

\begin{figure}[b]
\includegraphics[width=0.48\columnwidth]{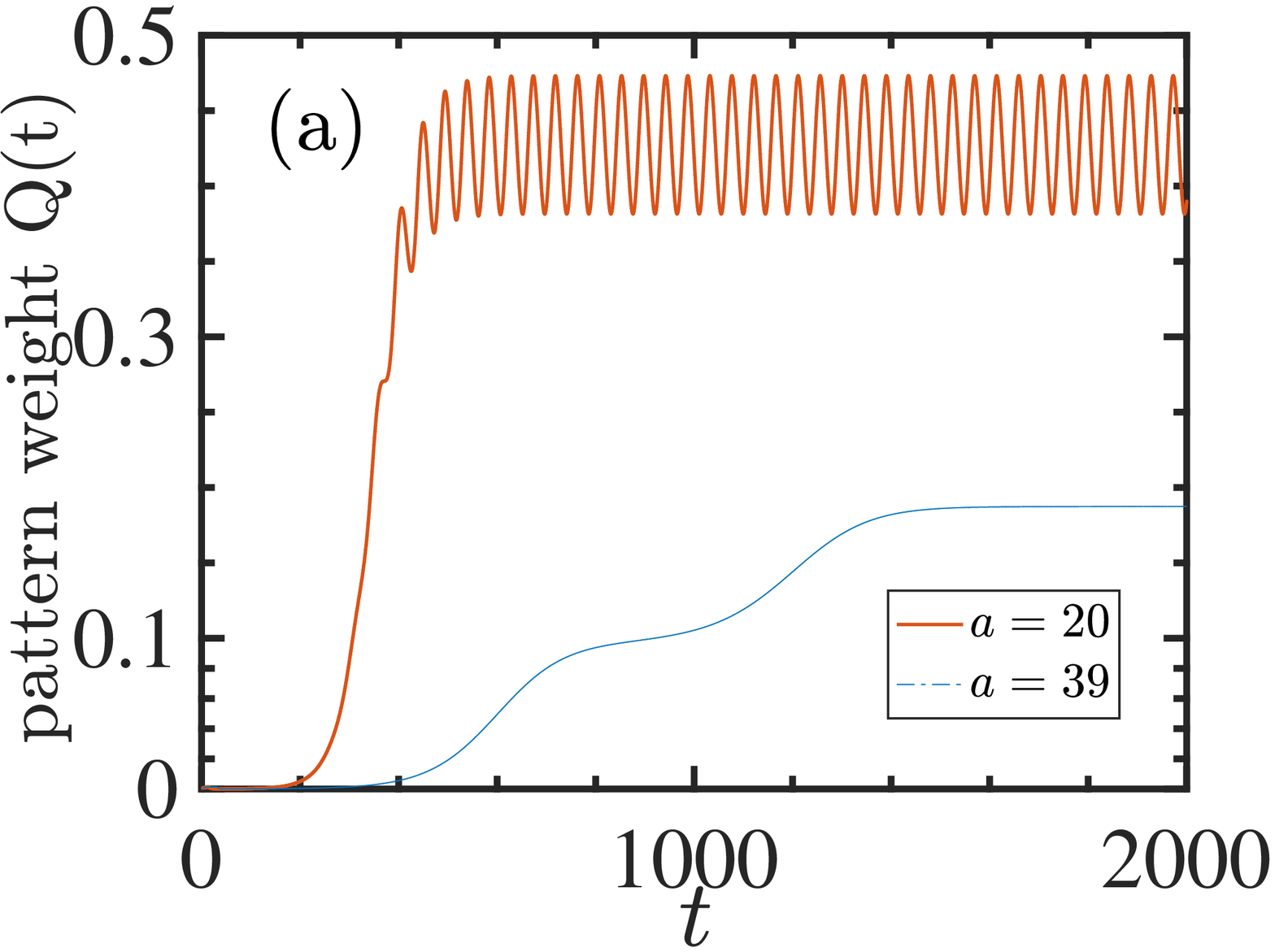}
\includegraphics[width=0.48\columnwidth]{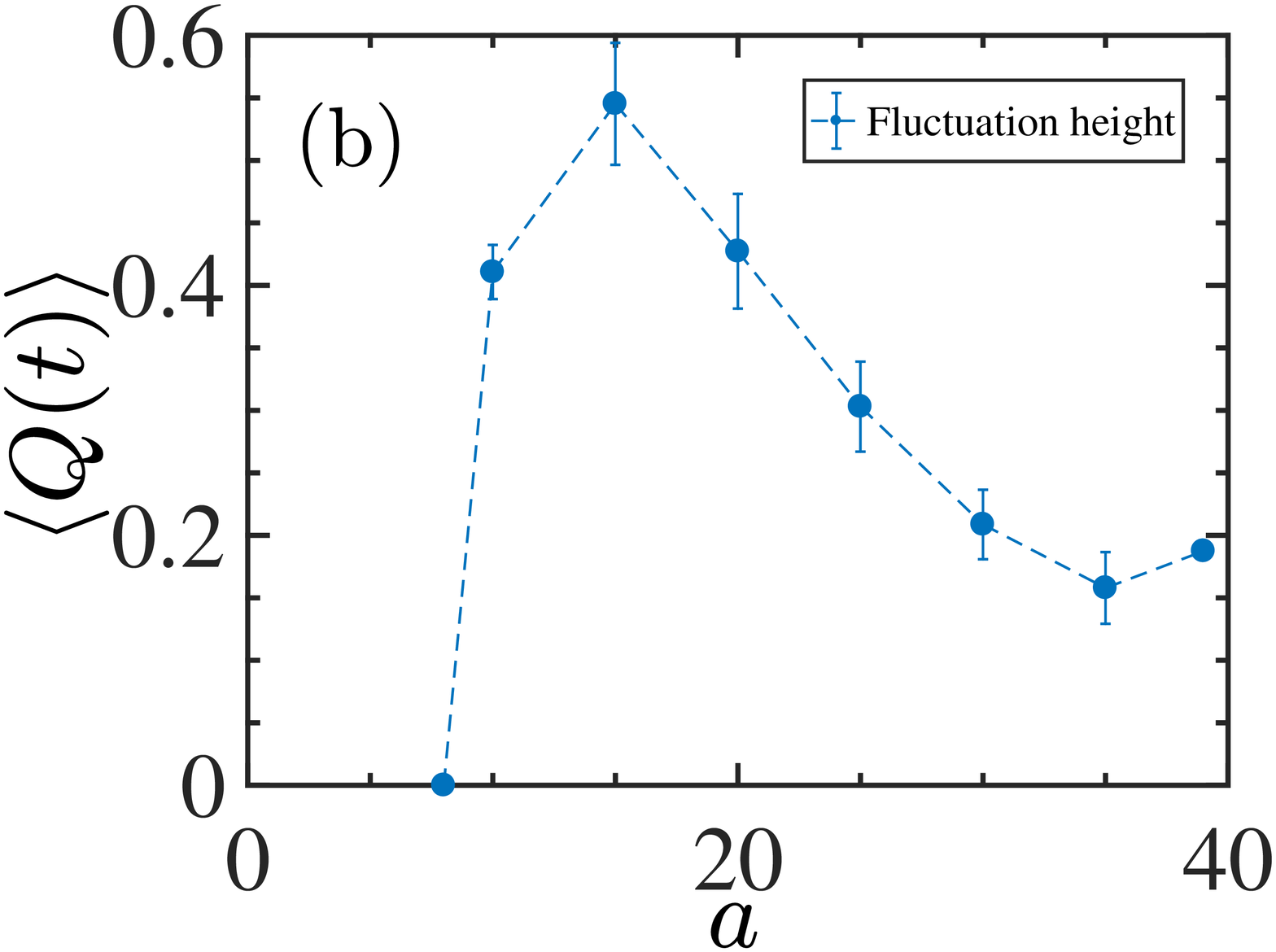}
	\caption{(a) Typical dependence of the pattern weight $Q(t)$ on drive separation $a$. At large separations, $a=39$, two steady-state patterns emerge and the weight saturates to a constant value. At smaller separations, $a=20$, the pattern weight oscillates with the frequency $T$ of Rabi oscillations. (b) The (temporally) average pattern weight $\bar{Q}$ and the peak-to-peak fluctuation amplitude as a function of drive separation $a$ emergence of a Rabi-oscillating pattern ($\bar{Q}>0$) at $a\gtrsim 9$ and two steady-state patterns, with zero peak-to-peak fluctuation-amplitude, for $a\gtrsim 39$. System parameters are $\mu=0.1,\nu=0.32,\gamma_{0}=0.28$, and $\sigma=3$.}
	\label{Fig:Spatiotemporal_Energy_a20_sigma3}
\end{figure}

Figure~\ref{Fig:Spatiotemporal_Energy_a20_sigma3}a shows that when the drives are well separated, $a=39$, the weight $Q(t)$ becomes constant after the random initial fluctuations amplify into a steady-state pattern. On the other hand, when the pattern undergoes Rabi oscillations at $a=20$, the pattern weight $Q(t)$ also oscillates around a steady-state mean value. Figure~\ref{Fig:Spatiotemporal_Energy_a20_sigma3}b shows the dependence of the average pattern weight $\bar{Q}$ and its peak-to-peak oscillation amplitude as a function of $a$. At small drive separation, both are zero as there is no pattern. At intermediate values of $a$, due to the Rabi oscillations, both are nonzero. At large separations, the oscillation amplitude goes to zero as Rabi oscillations cease and two steady-state patterns emerge. 

These results, too, are reminiscent of the non-unitary energy dynamics in a $\mathcal{PT}$-symmetric dimer with a saturating gain. We will, therefore, call the transition from a Rabi oscillating pattern to two steady-state patterns a $\mathcal{PT}$ symmetry breaking transition, although the system described by Eq.(\ref{Eq:PPDNLS-1}) has a global dissipative term and is not invariant under combined parity and time-reversal operations. 

\begin{figure}[h]
\centering
	\includegraphics[width=0.5\columnwidth]{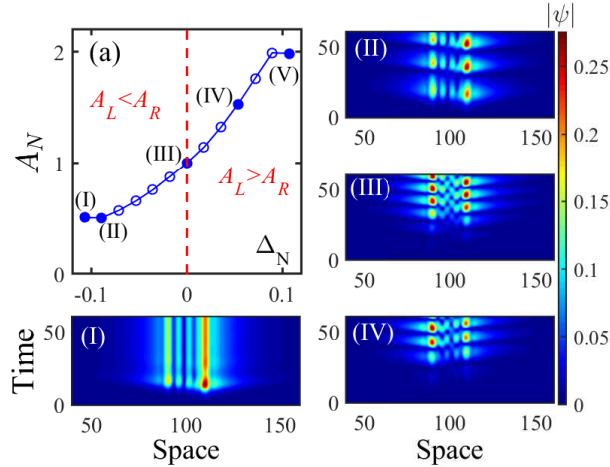}
	\caption{The values of parameters are  $\varphi=\pi,  \mu=0.1, \nu=0.32, \gamma_{0}=0.28$, $\sigma=3$, $a=20$ and $\delta\in[-0.03,0.03]$. (a) shows $A_{N}=A_{L}/A_{R}$ as a function of effective mismatch $\Delta_{N}$. The vertical dashed line (in $\Delta_{N}=0$) separates two regions, the left region is when $A_{L}<A_{R}$ $(\gamma_{L}<\gamma_{R})$ and the right region is when $A_{L}>A_{R}$ $(\gamma_{L}>\gamma_{R})$. The Figs.~\textrm{I}-\textrm{IV} are spatiotemporal diagrams that correspond to the marked points in panel (a).}
	\label{Fig:Mismatch_spatiotemporal}
\end{figure} 
\noindent{\it Robustness of the pattern dynamics.} In traditional, linear $\mathcal{PT}$ system, the $\mathcal{PT}$ symmetry breaking transition is highly susceptible to asymmetric perturbations that immediately render the eigenvalues complex; indeed, this susceptibility is characterized by the degree of the exceptional point at which the $\mathcal{PT}$ transition occurs. Contrarily, we will now show that the $\mathcal{PT}$ (breaking or restoring) transition shown by the pdNLSe is robust with respect to the such perturbations. 

To this end, we consider two localized, out-of-phase drives with different amplitudes $\gamma_0\pm\delta$, and investigate the dependence of the resulting, stable, pattern amplitudes $A_L$ and $A_R$ on the mismatch. Figure~\ref{Fig:Mismatch_spatiotemporal}a shows the normalized amplitude $A_N=A_L/A_R$ as a function of the normalized mismatch strength $\Delta_N(\delta)=\delta/\gamma_0$. When $\Delta_N<0$, the drive on the left is smaller in strength than the drive on the right, and we get $A_N<1$.  On the other hand, when $\delta>0$, the larger local drive on the left is reflected in the larger normalized amplitude $A_N>1$. Corresponding spatiotemporal maps for $|\psi(x,t)|$ are shown in panels I-IV. At $\Delta_N=-0.11$ (panel I) the system is in the $\mathcal{PT}$-symmetry broken phase, and the steady-state pattern on the right has a larger amplitude, consistent with its larger parametric drive strength. When the mismatch is reduced to $\Delta_N=0.08$ (panel II), the system develops Rabi oscillations, but with larger amplitude on the right and smaller one on the left. As the mismatch changes from negative to positive, this asymmetry is reversed, and the amplitude of the oscillating pattern is largest on the left hand side (panels III, IV). 

These results show that the Rabi oscillations induced by two, out-of-phase, localized parametric drives are robust with respect to the mismatch and this mismatch can be used to manipulate the size and the dynamics of resulting localized patterns.

\begin{figure}[t]
\includegraphics[width=0.6\columnwidth]{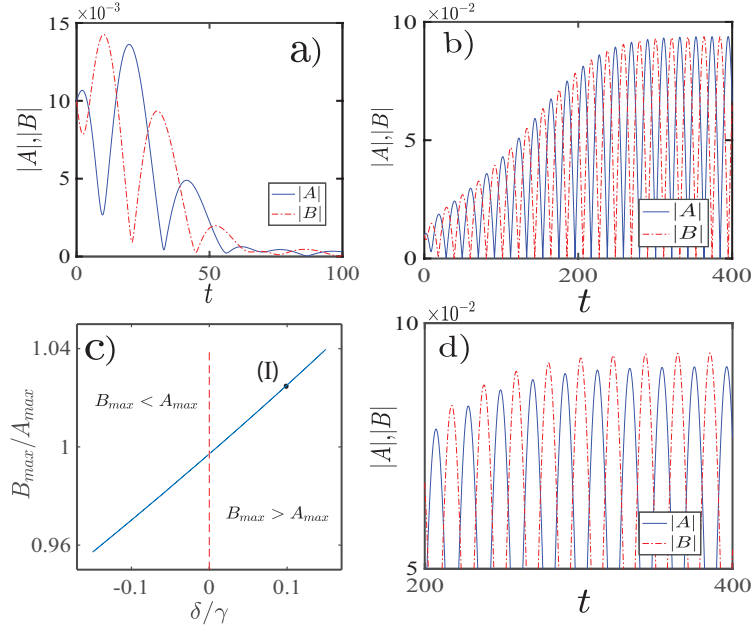}
\caption{Phenomenology of a parametric dimer with $\mu=0.05,\nu=0.23$, and $\gamma=0.1$. (a)  when $\epsilon=0.12$, fluctuations decay and there is no pattern. (b) for $\epsilon=0.15$, an oscillating pattern with equal amplitude on the two sites emerges. (c) ratio of amplitudes $B_\mathrm{max}/A_\mathrm{max}$ as a function of drive mismatch $\delta/\gamma$ shows the expected behavior. (d) $|A(t)|$ and $|B(t)|$ corresponding to point marked (I) in panel c.}
\label{Fig:Mismatch_temporal}
\end{figure}

\noindent {\it A parametric dimer model.} To elucidate some results of the continuum system, we consider a two-site discretization~\cite{Barashenkov2014} of Eq.(\ref{Eq:PPDNLS-1}), i.e.  a dimer model with nonlinearity, friction, and site-dependent parametric driving, 
\begin{eqnarray}
\frac{dA}{dt} &=& -(\mu+i\tilde{\nu}+i\left\vert A\right\vert^{2}) A-i\epsilon B +\gamma A^*,
\label{eqn:mws1} \\
\frac{dB}{dt} &=& -(\mu+i \tilde{\nu}+i \left\vert B\right\vert^{2})B-i\epsilon A-\gamma B^*.
\label{eqn:mws2}
\end{eqnarray}
where $A$ and $B$ represent the complex amplitudes on the two sites and $\tilde{\nu}=\nu-\epsilon$. The decay or formation of a localized pattern and its Rabi-like oscillations are informed by a linear stability analysis of these equations around the equilibrium quiescent state $A=0+i0=B$. The resulting complex eigenvalues of the Jacobian matrix are given by 
\begin{eqnarray}
\label{eq:ev1}
\lambda_{1,3}& = & -\mu +\sqrt{\gamma^{2}-(\epsilon-\nu)^{2}}\pm i\epsilon,\\
\lambda_{2,4}& = & -\mu-\sqrt{\gamma^{2}-(\epsilon-\nu)^{2}}\pm i\epsilon.
\end{eqnarray}
When $\epsilon=0$, there is no cross-talk between the two sites. If the discriminant $D(\epsilon)=\gamma^2-(\nu-\epsilon)^2$ is negative, the four eigenvalues $\lambda_k$ have identical dissipative real part, indicating that there is no pattern and fluctuations return the system to the quiescent state. When $D$ is positive, modes with eigenvalues $\lambda_{2,4}$ continue to decay, whereas the modes with eigenvalues $\lambda_{1,3}$ amplify when $\gamma>\sqrt{\nu^2+\mu^2}$, leading to the formation of a steady-state pattern from random initial fluctuations. When the two sites are coupled, $\epsilon\neq 0$, similar considerations imply that there is no pattern, i.e. all eigenvalues have negative real parts and all eigenmodes are dissipative, when $\epsilon<\epsilon_{-}\equiv\nu-\sqrt{\gamma^2-\mu^2}$ or $\nu+\sqrt{\gamma^2-\mu^2}\equiv\epsilon_{+}<\epsilon$. On the other hand, when $\epsilon_{-}\leq\epsilon\leq\epsilon_{+}$, the system has two amplifying eigenmodes, leading to a pattern that has Rabi oscillations. Figure~\ref{Fig:Mismatch_temporal}a-b shows the time-evolution of $A(t)$ and $B(t)$ obtained by solving Eqs.(\ref{eqn:mws1})-(\ref{eqn:mws2}) with infinitesimal, random initial conditions. 

This system is also robust against mismatch in drive strengths, $\gamma\rightarrow\gamma\pm\delta$. Figure ~\ref{Fig:Mismatch_temporal}c shows the dependence of the amplitude ratio $B_\mathrm{max}/A_\mathrm{max}$ on the normalized mismatch $\delta/\gamma$, and Fig.~\ref{Fig:Mismatch_temporal}d shows the corresponding time evolution. The qualitative similarities between results for the dimer and continuum models show that the simple model captures the salient features such as the presence of robust Rabi oscillations of dissipative structures. 

\noindent{\it Discussion:} Over the past decade, the ideas of $\mathcal{PT}$ symmetry and its breaking have been extensively explored in both linear and nonlinear settings. Here we have shown that dissipative structures, which are a quintessential nonlinear phenomena, exhibit dynamics that are very similar to that of a prototypical $\mathcal{PT}$-symmetric system undergoing a $\mathcal{PT}$-symmetry breaking transition. Remarkably, our model has a global frictional loss and out-of-phase gain drives, while a traditional $\mathcal{PT}$-symmetric system has balanced loss and gain. 

The Rabi oscillations of the resultant localized patterns, with a period that diverges at the transition, are induced by out-of-phase parametric drives that need not be balanced. This is of particular importance in experimental realization of such a system, say, in a shallow tank of water~\cite{Urra2017}. These features of our results for the continuum equation are captured by a parametric dimer model with friction, nonlinearity, and out-of-phase parametric drives. The emergence of a stable nonzero solution (dissipative structure) and the robustness of subsequent Rabi oscillations implies that this system provides a new way to ``balance gain and loss'' in the presence of global friction and multiple parametric gains. 


M.A.G-\~N. acknowledges financial support of Proyecto Puente-PUCV. F.R.M-H thanks to Universidad de Tarapac\'a for the financial support project N$^{\circ}$ 4738-19. Y.N.J. was supported by NSF-DMR 1054020. 

%

\end{document}